\documentclass{epl}
\usepackage{graphicx}

\newcommand{\Ca} {{\rm Ca}}
\newcommand{\DeB} {{\rm De}_B}
\newcommand{\DeK} {{\rm De}_K}
\newcommand{\eg} {{{\it e.g.}, }}
\newcommand{\etaM} {\eta_{\rm M}}
\newcommand{\etaT} {\eta_{\rm T}}
\newcommand{\etaV} {\eta_{\rm V}}
\newcommand{\etaTV} {\eta_{\rm T,V}}
\newcommand{\etaW} {\eta_{\rm W}}

\newcommand{\ie} {{{\it i.e.}, }}
\newcommand{\pd} {\partial}
\newcommand{\R} {{\rm Re}}
\newcommand{\rmB} {{\rm B}}
\newcommand{\rmd} {{\rm d}}
\newcommand{\rmr} {{\rm r}}
\newcommand{\rms} {{\rm s}}

\newcommand{\vecr} {{\bf r}}

\newcommand{\vs} {v_{\rms,\alpha}}

\title{Surface relaxation of lyotropic lamellar phases}
\shorttitle{Surface relaxation of lamellar phases}
\author{H.\ Bary-Soroker\inst{1} \and H.\ Diamant\inst{2}}

\institute{
  \inst{1} School of Physics \& Astronomy\\
  \inst{2} School of Chemistry\\ Raymond \& Beverly Sackler Faculty of
  Exact Sciences, Tel Aviv University, Tel Aviv 69978, Israel }

\begin{document}

\maketitle

\begin{abstract}
  We study the relaxation modes of an interface between a lyotropic
  lamellar phase and a gas or a simple liquid. The response is found
  to be qualitatively different from those of both simple liquids and
  single-component smectic-A liquid crystals. At low rates it
  is governed by a non-inertial, diffusive mode whose decay rate
  increases quadratically with wavenumber, $|\omega|=Aq^2$. The
  coefficient $A$ depends on the restoring forces of surface tension,
  compressibility and bending, while the dissipation is dominated by
  the so-called slip mechanism, \ie relative motion of the two
  components of the phase parallel to the lamellae.  This surface mode
  has a large penetration depth which, for sterically stabilised
  phases, is of order $(dq^2)^{-1}$, where $d$ is the microscopic
  lamellar spacing.
\end{abstract}

\begin{figure}[tbh]
\centerline{\resizebox{0.47\textwidth}{!}
{\includegraphics{fig1a.eps}}
\hspace{0.3cm}
\resizebox{0.47\textwidth}{!}
{\includegraphics{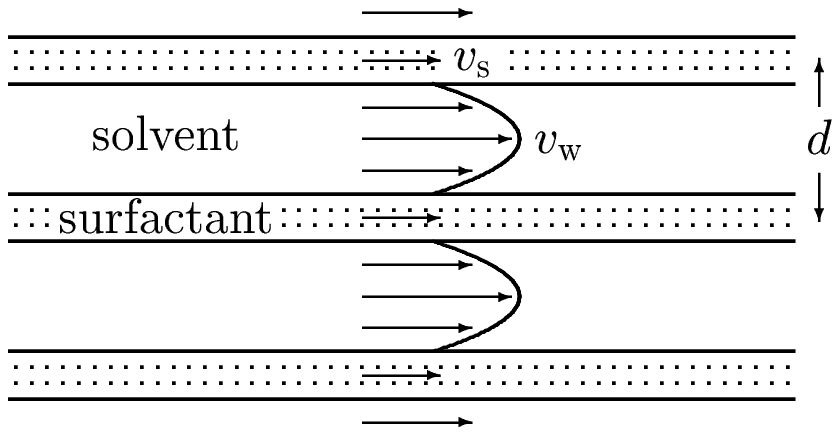}}}
\vspace{-4cm}
\hspace{0.06\textwidth}
a)
\hspace{0.42\textwidth}
b)
\vspace{4cm}
\caption[]{a) Schematic view of the system. 
A semi-infinite lamellar phase occupies the region $z<0$, having an
interface at $z=0$ with a gas or a simple liquid. A surface
perturbation is shown, whose amplitude decays to zero at $z\rightarrow
-\infty$.  b) Visualisation of the slip flow, in which the two
components of the lamellar phase have different average velocities.}
\label{fig1}
\end{figure}

Lyotropic lamellar phases occupy large portions in the phase diagrams
of amphiphilic molecules (surfactants) in solution
\cite{surfactant,Safran}.  They consist of stacks of parallel fluid 
membranes separated by microscopic ($\sim$1--10 nm thick) layers of
solvent [fig.\ \ref{fig1}(a)], thus having the symmetry of a smectic-A
liquid crystal \cite{LC}.  These phases appear in numerous
applications, \eg in the cosmetic and detergent industries. Lamellar
bodies are found also in biological systems such as the lung
\cite{lung}.
Apart from the free surfaces of lamellar phases with air, surfactant
phase diagrams contain also large coexistence regions (so-called
immiscibility gaps) \cite{surfactant}, in which a lamellar phase has
an equilibrium interface with an isotropic liquid such as a dilute
micellar solution or an $L_3$ (sponge) phase \cite{surfactant,Safran}.
Spherical lamellar structures in the form of multilamellar vesicles
(onions) dispersed in a solvent are commonly encountered as well
\cite{onion}.  Thus, the surface response of these phases is a
fundamental issue relevant to a large variety of experimental systems.
The static response of smectics to surface deformations was thoroughly
studied \cite{LC,Durand,Fournier}. The dynamics of surface
perturbations in thermotropic (single-component) smectics were
investigated as well, for both semi-infinite systems and finite films
\cite{Orsay,Holyst,Toner1,Toner2,Toner3,Chen,Romanov00,Romanov01,Romanov02}.
In the current Letter we analyse the surface relaxation of lyotropic
(two-component) lamellar phases and demonstrate the essentially
different surface dynamics of this ubiquitous class of materials.

Surface modes characterise the relaxation dynamics of surface
perturbations whose amplitude decays with increasing distance into the
bulk material [fig.\ \ref{fig1}(a)]. In a simple liquid, having mass
density $\rho$, viscosity $\eta$ and surface tension $\gamma$, the
surface dynamics depend on two dimensionless numbers: the Reynolds
number, $\R\equiv\rho|\omega|/(\eta q^2)$, and the Capillary number,
$\Ca\equiv\eta|\omega|/(\gamma|q|)$, where $q$ and $\omega$ are the
perturbation wavenumber and frequency, respectively. If inertia is
neglected, one is left with the Capillary number alone, leading to a
decay rate $|\omega|\sim\gamma|q|/\eta$.  However, substituting this
result back in the requirement $\R\ll 1$ yields
$|q|\gg\rho\gamma/\eta^2$. Thus, \eg for water such a non-inertial,
overdamped response is restricted to wavelengths much smaller than
$10^{-2}$ $\mu$m, while larger-wavelength perturbations are governed
by inertial capillary waves \cite{capillarywave}, whose dispersion
relation is set by $\R\Ca\sim 1$.

The viscoelastic hydrodynamics of smectic-A liquid crystals \cite{LC}
were formulated in the seminal work of ref.\ \cite{MPP}. Due to the
anisotropy of these materials the viscosity tensor contains five
independent coefficients which, for incompressible flows, can be
reduced to three, denoted by $\etaM$, $\etaT$ and $\etaV$
\cite{Brochard}. While $\etaM$ is related to relative sliding of layers, 
and is therefore comparable to the viscosity of the solvent (water) in
lyotropic phases, the other two are associated with distortions of the
membranes themselves and, hence, are comparable to the overall
apparent viscosity of the liquid, which is 2--3 orders of magnitude
larger \cite{Brochard,Roux2}.  The elasticity is characterised by a
compression modulus $B$ and a bending modulus $K$.  These elastic
moduli define two Deborah numbers in addition to the Capillary and
Reynolds numbers.  The dynamics thus depend on four dimensionless
numbers, which we define as
\begin{equation}
  \DeB \equiv \etaM|\omega|/B,\ \
  \DeK \equiv \etaM|\omega|/(Kq^2),\ \ 
  \Ca \equiv  \etaM|\omega|/(\gamma|q|),\ \ 
  \R  \equiv  \rho|\omega|/(\etaM q^2).
\label{numbers}
\end{equation}
(More accurately, in the case of lyotropic phases one should consider
separately, depending on the frequency, two layer-compression moduli
--- the modulus at constant surfactant concentration (for high
frequencies) and the modulus at constant surfactant chemical potential
(low frequencies), which is typically much smaller \cite{B}. Our main
focus here is on the slow response for which $B$ should be regarded,
unless otherwise noted, as the smaller, fixed-chemical-potential
modulus.)

The surface modes of a semi-infinite thermotropic smectic phase were
studied in ref.\ \cite{Romanov00}. Despite the much higher viscosity
of these phases compared to ordinary simple liquids (water), the
conclusion of this work was that the low-frequency surface modes were
inertial, elastic (Rayleigh) waves of second sound, their dispersion
relation being set by $\R\DeB\sim 1$. In a detailed analysis of
low-Reynolds-number surface modes of thermotropic smectics
\cite{next}, we have reached a similar conclusion, namely, that
non-inertial relaxation is restricted to small (sub-micron)
wavelengths. On the other hand, it is hard to envisage the surface
dynamics of a small lamellar droplet or an onion (of, say,
$10$--$10^2$ $\mu$m scale), whose inter-layer spacings and
surroundings are pervaded by viscous solvent, as dominated by
underdamped inertial waves. The disagreement between this intuition
and the aforementioned results suggests that lyotropic lamellar phases
may be dominated by another dissipative mechanism, different from the
usual viscous one.

The hydrodynamic theory of smectics was extended to lyotropic phases
in refs.\ \cite{Brochard,Nallet,Bruinsma}.  Brochard and de Gennes
\cite{Brochard} were the first to recognise that, 
unlike isotropic binary mixtures, the spatial organisation of the two
components in a lamellar phase allows for {\em collective} motion of
one component relative to the other parallel to the layers [fig.\
\ref{fig1}(b)]. This bulk slip mode was observed experimentally
\cite{Nallet,slip_exp1,slip_exp2,slip_exp3,slip_exp4}. We show in the
current work that the additional dissipation due to slip has profound
consequences for the surface dynamics. The dominance of this
dissipative mechanism can be demonstrated by the following simple
argument (to be made rigourous later on). The transport coefficient
associated with slip, $\mu$, relates to the mobility of a Poiseuille
flow of solvent between two membranes [fig.\
\ref{fig1}(b)], $\mu\sim d^2/\etaW$, where $d$ is the inter-membrane
spacing and $\etaW$ the solvent viscosity
\cite{Brochard}. The resulting friction force per unit volume is
$\sim\mu^{-1}v$, $v$ being a characteristic relative flow velocity,
while the force density arising from viscous stresses is $\sim\etaTV
q^2v$.  Hence, the slip dissipation dominates for $q\ll
(\etaTV\mu)^{-1/2} \sim(10d)^{-1}$ --- a condition satisfied for all
physically relevant wavelengths.

The system under consideration is schematically depicted in fig.\
1(a).  A semi-infinite lamellar phase occupies the region $z<0$, the
layers lying, on average, parallel to the $xy$ plane.  At $z=0$ there
is an interface between the lamellar phase and a dilute gas or a
simple liquid. We use Latin indices ($i,j$) to denote vector
components along all three axes $(x,y,z)$ and Greek indices
($\alpha,\beta$) for components along the two lateral directions
$(x,y)$.

The hydrodynamic description of thermotropic smectics requires six
scalar hydrodynamic fields \cite{LC,MPP}. Assuming an isothermal,
incompressible flow, one is left with only three independent fields,
\eg two of the three components of the flow velocity $v_i(\vecr,t)$ at
position $\vecr$ and time $t$, and an additional scalar field,
$u(\vecr,t)$, defining the local displacement of the layers from their
equilibrium position. Three dynamic modes result: two second-sound
waves and a transverse shear (vorticity) diffusive mode. In lyotropics
another scalar field is required \cite{Brochard,Nallet}, \eg the local
surfactant volume fraction. As a result, another mode emerges, \ie the
slip (baroclinic) mode.
In writing the hydrodynamic equations we use the variables
$v_i(\vecr,t)$, $u(\vecr,t)$, and a pressure field $p(\vecr,t)$ (of
which, we recall, only three are independent).  To include the slip
degree of freedom in lyotropics another velocity field is added
\cite{Brochard}, which we take as the lateral velocity of the
surfactant component, $\vs(\vecr,t)$.  Velocity differences between
the two components, $\vs-v_\alpha$, are driven by in-plane stresses in
the membranes
\cite{Brochard}. We focus here on the simplest treatment, 
where the membranes are considered as incompressible two-dimensional
(2D) liquids with no 2D viscous stresses. (This assumption is also
consistent with the limit, mentioned above, of the
fixed-chemical-potential compression modulus being much smaller than
the fixed-concentration one.) In this description the driving force
for slip is provided solely by a lateral pressure gradient,
$\pd_\alpha\psi(\vecr,t)$. Out of the three new variables, $\vs$ and
$\psi$, similarly, only one is independent.

In terms of the aforementioned variables, the hydrodynamic equations
for a lamellar phase take the form
\begin{eqnarray}
  && \rho(\pd_t v_i + v_j\pd_j v_i) = \pd_j\sigma_{ij} 
 \label{NS}\\
  && \pd_i v_i = 0
 \label{conserv}\\
  && \pd_\alpha v_{\rms,\alpha} = 0 
 \label{conserv_s}\\
  && v_{\rms,\alpha} - v_\alpha = -\mu \pd_\alpha \psi 
 \label{slip}\\
  && \pd_t u - v_z = 0.
 \label{permeation}
\end{eqnarray}
Equation (\ref{NS}) is the Navier-Stokes equation, where
$\sigma_{ij}$, to be specified below, is the stress tensor. Equations
(\ref{conserv}) and (\ref{conserv_s}) impose mass conservation on the
incompressible overall flow and 2D surfactant flow, respectively.
Equation (\ref{slip}) asserts a linear relation between the slip
velocity and its driving force. In eq.\ (\ref{permeation}) we have
assumed that permeation of material perpendicular to the layers is
negligible
\cite{LC}. We write the stress tensor as
$\sigma=\sigma^\rmr + \sigma^\rmd$, where the ``reactive'' part is
given by \cite{LL_elast,Brochard}
\begin{equation}
  \sigma^\rmr_{xx} = \sigma^\rmr_{yy} = -p - \psi + K\pd_{\beta\beta z}u,\ \ 
  \sigma^\rmr_{zz} = -p + B\pd_z u,\ \ 
  \sigma^\rmr_{\alpha z} = -K \pd_{\beta\beta\alpha} u,\ \ 
  \sigma^\rmr_{xy} = 0,
\end{equation}
and the dissipative part by \cite{Brochard}
\begin{equation}
  \sigma^\rmd_{\alpha\beta} = \etaT(\pd_\alpha v_\beta + \pd_\beta v_\alpha),\ \ 
  \sigma^\rmd_{zz} = 2\etaV\pd_z v_z,\ \ 
  \sigma^\rmd_{\alpha z} = \etaM(\pd_\alpha v_z + \pd_z v_\alpha).
\end{equation}

At $z=0$ the tangent force per unit area, $\sigma_{\alpha z}$, must
change continuously across the interface, whereas the perpendicular
one, $\sigma_{zz}$, has a discontinuity due to surface tension. If the
surface is free (the other phase is a dilute gas), these conditions
reduce to
\begin{equation}
  z=0:\ \ \ \sigma_{\alpha z} = 0,\ \ \ 
  \sigma_{zz} = \gamma \pd_{\alpha\alpha} u.
\label{BC1}
\end{equation}
Since we deal with surface perturbations, the bulk far away from the interface is
assumed to remain at equilibrium,
\begin{equation}
  z\rightarrow -\infty:\ \ \  v_i = \vs = u = 0.
\label{BC2}
\end{equation}
If the phase in the region $z>0$ is a liquid, its hydrodynamic
equations should be added, and the boundary conditions at $z=0$ are
modified. Since the introduction of a simple liquid leads to a minor
effect \cite{next}, we shall briefly comment on it below and
otherwise focus on the simpler case of a free surface.

We are concerned with small deviations from equilibrium and,
therefore, the nonlinear term in eq.\ (\ref{NS}) can be neglected.
Equations (\ref{NS})--(\ref{BC2}) then define a closed set of linear
equations with the appropriate boundary conditions, from which the
surface modes can be derived. For simplicity we restrict the
discussion to variations in the $x$ and $z$ directions only.
Substituting a perturbation of the form $ f(x,z,t)\propto
e^{iqx-i\omega t} $, where $f$ stands for any of the hydrodynamic
variables, we obtain a 4th-order equation for $v_z$,
\begin{equation}
  \pd_{zzzz}v_z - (S + \Theta - 2 - \DeB^{-1} - 
  \R ) q^2 \pd_{zz} v_z + (1 - \DeK^{-1} - \R )q^4v_z = 0.
\label{vz1}
\end{equation}
In eq.\ (\ref{vz1}) the dynamic numbers ($\DeB,\DeK,\R$) are as
defined in eq.\ (\ref{numbers}) with $|\omega|$ replaced by $i\omega$,
and $S$ and $\Theta$ are two additional dimensionless parameters, the
former associated with slip, $S\equiv (\etaM\mu q^2)^{-1}$, and the
latter with the viscosity anisotropy, $\Theta\equiv
2(\etaT+\etaV)/\etaM\gg 1$. The solution to eq.\ (\ref{vz1}) is given
by
\begin{eqnarray}
  v_z &=& (C_+ e^{\alpha_+ z} + C_- e^{\alpha_- z}) e^{iqx-i\omega t},
\ \ \ 
  \alpha_\pm = |q| [(S + \Theta - 2 - \DeB^{-1} - \R \pm \Gamma^{1/2})/2]^{1/2}
 \nonumber\\
  \Gamma &\equiv& (S + \Theta - 2 - \DeB^{-1} - \R)^2 - 4(1 - \DeK^{-1} - \R),
\label{vz2}
\end{eqnarray}
where the two spatial decay coefficients, $\alpha_\pm$, have been
chosen as the ones with a positive real part so as to satisfy the
boundary conditions at $z\rightarrow -\infty$, eq.\ (\ref{BC2}).
Imposing the interfacial boundary conditions (\ref{BC1}) yields two
linear equations for the amplitudes $C_+$ and $C_-$, whose determinant
is set to zero to find the dispersion relation,
$\omega=\omega(q)$. The resulting equation is
\begin{equation}
  \Ca^{-1}(\alpha_+ + \alpha_-)/|q| + \R - (1-\DeK^{-1}-\R)^{1/2}
  (S + \Theta - \DeB^{-1} - \DeK^{-1} - \R) = 0,
\label{det}
\end{equation}
where $\Ca$ is as defined in eq.\ (\ref{numbers}) with the
replacement $|\omega|\rightarrow i\omega$.

Equation (\ref{det}), together with the expressions for $\alpha_\pm$
given in eq.\ (\ref{vz2}), can be solved for $\omega$ only
numerically.  However, we can significantly simplify the problem by
noticing the following.  First, the slip mechanism will dominate over the
viscous one when $S\gg\Theta$. As already discussed above, this
amounts to a condition for $q$ which is practically always valid.
Secondly, for typical lamellar phases the length $\lambda=(K/B)^{1/2}$
is very small (of order a few $d$) and, as a result,
$\DeB^{-1}\gg\DeK^{-1}$ for any relevant $q$. We also take
$\R\ll\DeK^{-1}<1$, $\Theta\ll\DeB^{-1}$, and $S>\DeB^{-1}$.  (The
self-consistency of these conditions will be later verified.)  Under
these assumptions we find $\alpha_+/|q| \simeq (S-\DeB^{-1})^{1/2} \gg
\alpha_-/|q| \simeq [(1-\DeK^{-1})/(S-\DeB^{-1})]^{1/2}$.
Consequently, eq.\ (\ref{det}) is simplified to
\begin{equation}
  \Ca^{-1} - [(1-\DeK^{-1})(S - \DeB^{-1})]^{1/2} = 0,
\label{det2}
\end{equation}
whose solution is
\begin{equation}
  \omega = -i A q^2,\ \ \ 
  A = [B\mu + K/\etaM + \sqrt{(B\mu - K/\etaM)^2 + 4\mu\gamma^2/\etaM}]/2.
\label{omega}
\end{equation}

Equation (\ref{omega}) is our central result. It describes a decay
rate of a diffusive mode, which is affected by all three restoring
forces and the slip dissipative mechanism. To make these results more
transparent, let us specialise to sterically stabilised lamellar
phases, where $B\sim (k_\rmB T)^2/(\kappa d^3)$, $K\sim\kappa/d$, and
$\gamma\sim k_\rmB T/d^2$, $k_\rmB T$ being the thermal energy and
$\kappa$ the bending modulus of a single membrane, typically a few
$k_\rmB T$.  (The low tension values in lyotropic phases, $\gamma\sim
k_\rmB T/d^2$, arise from the diffusive contact of the surface
membranes with membranes in the bulk, which are essentially
tensionless.)  Recalling that $\etaM\mu\sim d^2$, we find for such
phases that all the terms appearing in eq.\ (\ref{omega}) are of the
same order and $A\sim k_\rmB T/(\etaM d)$. This yields $10^2$--$10^3$
$\mu$m$^2$/s for $d=10$ nm and $\etaM=10^{-2}$ poise.  The spatial
decay coefficients in this case are $\alpha_+\sim d^{-1}$ and
$\alpha_-\sim dq^2$. Thus, the surface mode contains a boundary layer
of microscopic thickness $\sim d$, and a deeply penetrating undulation
whose thickness is much larger than the wavelength.  The amplitude
ratio of these two parts is $C_+/C_-\simeq -\alpha_-^2/q^2$, which is
a small (negative) number, \ie the dominant part is the deeply
penetrating one.

Upon substitution of the results in the initial assumptions, their
validity is readily confirmed.  In particular, the Reynolds number is
$\R\sim\rho k_\rmB T/(\etaM^2 d)\sim 10^{-4}$--$10^{-3}$ for the above
parameters and $\rho=1$ g/cm$^3$, {\em irrespective of $q$}.  In the
case of stiffer (electrostatically stabilised) phases where $B\gg
(k_\rmB T)^2/(\kappa d^3)$, however, the assumption $\R\ll\DeK^{-1}$
is satisfied only if $B\ll (K/\rho)^{1/2}/\mu$. For typical parameters
this implies $B\ll 10^9$ erg/cm$^3$. Hence, for very stiff lamellar
phases inertia may become important, and one should return to the more
general equations (\ref{vz2}) and (\ref{det}).

Static surface deformations of smectics have a large penetration depth
as well, $(\lambda q^2)^{-1}\sim (dq^2)^{-1}$ \cite{Durand,Fournier}.
Thus, the static and dynamic penetrations turn out to be similarly
deep despite their different physical origins. While in the static
case the large depth arises from the fact that the layers are much
more resistant to compression than to bending, in the dynamic case it
stems from the high resistance to relative motion of the two
components compared with the viscous resistance to their overall flow.
Once a deeply penetrating undulation exists, a thin boundary layer
must appear as well to satisfy the surface boundary condition for the
transverse stress, $\sigma_{\alpha z}=0$. If the boundary layer were
absent, the ratio between the two transverse-stress terms would be
$\pd_\alpha v_z/\pd_z v_\alpha \sim q^2/\alpha_-^2 \gg 1$, whereas in
the presence of a boundary layer of thickness $\alpha_+^{-1}$ we have
$\pd_\alpha v_z/\pd_z v_\alpha \sim (q^2/\alpha_+^2)(C_-/C_+) \sim
q^4/(\alpha_+\alpha_-)^2 \sim 1$, thus allowing the transverse-stress
contributions to balance one another.

It should be noted that, having employed the assumptions of
2D-incompressible membranes and negligible $\R$, we have suppressed
the second-sound and vorticity-diffusion modes, leaving only one bulk
mode (the slip, baroclinic one) of the four existing in bulk lyotropic
phases. Rayleigh waves, therefore, cannot emerge from this
analysis. Such modes should exist in lyotropics as in thermotropics
\cite{Romanov00}. Their frequency domain, however, is much higher than
that of the relaxation discussed here. For the two rates to be
comparable, one must have $(B/\rho)^{1/2}q\sim Aq^2$, \ie
$q^{-1}\sim(B\rho)^{1/2}\mu\sim 1$--$10$ nm. Hence, the time scales
are well separated for any relevant $q$ and the two responses can
safely be studied as decoupled.  (The separation of time scales is
even larger, in fact, since one should use for the fast second-sound
response the much larger compression modulus at constant concentration
\cite{B}.)

If the phase in the region $z>0$ is taken to be a simple liquid
\cite{next}, the only change is that the term $\Ca^{-1}$ in eqs.\ 
(\ref{det}) and (\ref{det2}) is replaced with
$(\Ca^{-1}-2\eta/\etaM)$, $\eta$ being the viscosity of the simple
liquid. Therefore, the simple liquid does not affect the above results
so long as $\eta\ll\gamma/(A|q|)$. For sterically stabilised lamellar
phases, for example, this implies that only when the simple liquid is
orders of magnitude more viscous than the solvent,
$\eta\sim\etaM/(d|q|)\gg\etaM$, will it have an appreciable effect on
the surface relaxation. This is yet another consequence of the
dominance of slip dissipation.

To summarise, in contrast with simple liquids and thermotropic
smectics, whose surface dynamics are governed by underdamped waves
(capillary or Rayleigh waves, respectively), we have found that
surfaces of lyotropic lamellar phases can relax via a much slower,
overdamped diffusive mode over a wide range of wavelengths. The key
ingredient underlying this qualitatively different behaviour is the
presence of viscous solvent in between the membranes and the resulting
relative slip of surfactant and solvent layers. The relaxation is
remarkably slow due to the strong friction (small $\mu$) introduced by
this motion. For example, in sterically stabilised phases
perturbations of a micron-scale wavelength are predicted to decay with
a rate of $\sim 10^3$ s$^{-1}$.  To our best knowledge, experiments
concerning the surface modes of lamellar phases have not yet been
performed. The predictions of this work should be verifiable in, \eg
dynamic scattering experiments.

In this study we have considered semi-infinite, flat lamellar phases.
Our results should be valid for finite films and curved lamellar
structures (\eg onions) provided that their size is sufficiently
large, or the wavelength is sufficiently small. This restriction,
however, is particularly severe in the current case because of the
deep penetration of the inferred surface mode. The onion radius, for
example, should be much larger than $\alpha_-^{-1}\sim (dq^2)^{-1}\gg
q^{-1}$. For a typical onion of radius $R\sim 10$ $\mu$m and $d\sim 1$
nm the wavelength must be much smaller than $(Rd)^{1/2}\sim 10^{-1}$
$\mu$m.  An extension of the theory to finite films and curved
surfaces is needed, therefore, to accurately account for the surface
dynamics of such lamellar objects.

Finally, lamellar surfaces {\em out} of equilibrium have been long
known to exhibit an intriguing instability involving multilayer
finger-like structures (myelin figures) \cite{myelin,myelinUC}.  We
hope that the elucidation of the relaxation of lamellar surfaces
presented here will be instrumental also in the resolution of this
long-standing puzzle.

\acknowledgments 

We thank D.\ Andelman, L.\ Bary-Soroker, M.\ Cates, B.\ Davidovitch,
S.\ Egelhaaf, and M.\ Kozlov for helpful discussions. This work was
supported by the US--Israel Binational Science Foundation (2002271).
H.D.\ acknowledges additional support from the Israeli Council of
Higher Education (Alon Fellowship).

\end{document}